\documentclass[aps,prx,twocolumn,groupedaddress,superscriptaddress,showpacs,amssymb,amsmath,amsfonts,floatfix,nofootinbib]{revtex4-2}

\usepackage{mathtools}
\usepackage{perpage}
\MakePerPage{footnote}
\usepackage{color}
\usepackage{braket}
\usepackage{subfig}
\usepackage{graphicx}
\usepackage{float}
\usepackage[unicode=true,bookmarks=true,bookmarksnumbered=false,bookmarksopen=false,breaklinks=false,pdfborder={0 0 1},backref=false,colorlinks=true]{hyperref}
\hypersetup{linkcolor=magenta, urlcolor=blue, citecolor=blue, pdfstartview={FitH}, hyperfootnotes=false, unicode=true}
\usepackage{booktabs}
\usepackage{MnSymbol}
\begin{document}

\title{Learning Hamiltonians for $O(1)$ Oracle-Query Quantum State Preparation}

\author{Sadegh Salami}
\affiliation{Centre for Quantum Engineering and Photonics Technology, Sharif University of Technology, Tehran, Iran}
\affiliation{Department of Computer Engineering, Sharif University of Technology, Tehran, Iran}

\author{Mohammad T. Manzuri}
\affiliation{Department of Computer Engineering, Sharif University of Technology, Tehran, Iran}

\author{Mehdi Ramezani}
\affiliation{Centre for Quantum Engineering and Photonics Technology, Sharif University of Technology, Tehran, Iran}

\author{Sina Asadiyan Zargar}
\affiliation{Centre for Quantum Engineering and Photonics Technology, Sharif University of Technology, Tehran, Iran}
\affiliation{Department of Electrical Engineering, Sharif University of Technology, Tehran, Iran}

\author{Abolfazl Bahrampour}
\affiliation{Centre for Quantum Engineering and Photonics Technology, Sharif University of Technology, Tehran, Iran}

\author{Alireza Bahrampour}
\affiliation{Centre for Quantum Engineering and Photonics Technology, Sharif University of Technology, Tehran, Iran}
\affiliation{Department of Physics, Sharif University of Technology, Tehran, Iran}

\date{\today}

\begin{abstract}
We propose a Hamiltonian-based quantum state preparation method implemented via a shallow parametrized quantum circuit. The approach learns the parameters of a diagonal Hamiltonian through a classical training phase, while the quantum circuit itself performs only fixed-depth Hamiltonian evolution and mixing operations. With oracle access to the learned Hamiltonian parameters, $N$ classical data values can be encoded into $n=\log_2{N}$ qubits using $O(1)$ quantum queries, shifting the overall computational cost to an $O(N\log{N})$ classical preprocessing stage. For structured datasets generated by an underlying function, oracle access can be avoided by expressing the Hamiltonian in the Walsh basis and retaining only a polynomial number of significant terms. In this regime, quantum state preparation is achieved in $\text{poly}(n)$ time using $\text{poly}(n)$ parameters, reaching infidelities on the order of $10^{-5}$. By restricting the Hamiltonian to one-local and two-local terms, the method naturally yields hardware-efficient circuits suitable for near-term quantum devices. This separation keeps the quantum execution shallow while moving optimization to classical computation.
\end{abstract}

\maketitle

\section{Introduction}
\label{sec:introduction}
The Quantum State Preparation (QSP) problem concerns efficiently encoding a classical dataset into a quantum state. Given a normalized dataset
\begin{equation}
	\textbf{x} = \{x_0,x_1,\dots,x_{N-1}\},
\end{equation}
where $\sum_{j=0}^{N-1} x_j^2 = 1$, the goal is to prepare the quantum state
\begin{equation}
	\ket{\psi} = \sum_{j=0}^{N-1} x_j \ket{j},
\end{equation}
where $\ket{j}$ denotes the computational basis state corresponding to the binary representation of index $j$. The state $\ket{\psi}$ lives in the Hilbert space of $n=\log_{2}N$ qubits. 

QSP serves as a fundamental subroutine across a wide range of quantum algorithms. Amplitude-encoded states enable efficient implementation of quantum linear-algebra routines, including the Harrow-Hassidim-Lloyd (HHL) algorithm \cite{harrow2009quantum} for solving linear systems, quantum least-squares regression, and matrix inversion methods used in scientific computing and machine-learning pipelines. Many quantum machine-learning models---such as quantum kernel methods, variational classifiers, and quantum generative models---require loading classical data into quantum registers as an initial step \cite{schuld2015introduction, cerezo2022challenges,ramezani2025parallel}. QSP also underlies quantum Monte Carlo techniques \cite{montanaro2016quantum}, amplitude estimation \cite{brassard2000quantum,suzuki2020amplitude}, and probability-distribution loading used in quantum finance \cite{egger2020quantum,rebentrost2018quantum}, Bayesian inference \cite{cenxin2023quantum}, and quantum simulation of physical systems \cite{georgescu2014quantum,mcardle2020quantum}. Because these applications all rely on accurate and efficient preparation of amplitude-encoded states, improving the time complexity of QSP has direct implications for the practical feasibility of quantum algorithms on near-term hardware.

In general, deterministic algorithms for constructing $\ket{\psi}$ from arbitrary data require time complexity $O(N)$ with space complexity $O(n)$ \cite{GroverRudolph02,Mottonen05,Bergholm05,Shende06,Plesch11}. However, this cost can be reduced to $O(n)$ if one allows the use of $O(N)$ ancillary qubits \cite{Araujo21,Sun23,Gui24,nmaju2025low}. 

In many situations, the target dataset is not fully dense but exhibits additional structure---for example, when only a small subset of amplitudes is nonzero. Such sparsity effectively reduces the dimension of the state that must be synthesized, allowing the preparation cost to scale with the size of the support rather than the full Hilbert-space dimension. Sparse amplitude profiles therefore constitute an important class of structured datasets for which significantly fewer computational resources may be sufficient, and several dedicated sparse-QSP frameworks exploit this advantage \cite{Gleinig21,Malvetti21,Mozafari22DD,LiLuo25}.

Alternatively, if one has quantum oracle access to a function that efficiently generates the data values $x_i$, the time complexity can be reduced to $O(\sqrt{N})$ using amplitude amplification techniques \cite{grover2000synthesis,sanders2019black,bausch2022fast}.

In many practical scenarios, the dataset represents either the sampled values of a function or data drawn from a specific probability distribution. In such cases, employing truncated Fourier or Walsh series expansions can significantly reduce the time complexity of quantum state preparation \cite{moosa2023linear, zylberman2024efficient,boosari2025hybrid}.

A more general framework for efficient state preparation is provided by the Quantum Singular Value Transformation (QSVT) formalism. In this approach, one begins by constructing a block-encoding of a diagonal operator $A=\text{diag}(y_0,y_1,\dots,y_{N-1})$ whose entries $y_i$ can be generated efficiently, for example through a known analytic or functional dependence. If the desired amplitudes satisfy a relation $x_i=h(y_i)$ for some appropriately chosen polynomial or smooth function $h$, then QSVT enables the implementation of a new block-encoding that effectively realizes the operator $h(A)$. Applying this transformed operator to a reference state (such as a uniform superposition) produces the normalized quantum state $\ket{\psi}\propto\sum_{j}x_j\ket{j}$ \cite{mcardle2022quantum, li2023efficient, gonzalez2024efficient}. 

Another general approach to quantum state preparation employs variational learning of parameterized quantum circuits. In such methods, a quantum circuit composed of tunable single- and two-qubit gates is repeatedly executed while a classical routine optimizes its parameters to minimize a cost function that measures the distance between the circuit's output state and the desired target distribution. Through this iterative hybrid loop, the circuit learns to approximate the underlying function or probability density directly from data. This data-driven methodology dispenses with explicit analytical decomposition of amplitudes and can flexibly adapt to structured datasets, making it well suited for approximate preparation of states that are otherwise difficult to compute analytically~\cite{zoufal2019quantum,nakaji2022approximate,marin2023quantum}.

However, variational schemes intrinsically require large numbers of quantum evaluations per training epoch, so their overall runtime---and consequently, their practical scalability---is still constrained by the short decoherence times of present-day hardware.

In this work, we introduce a quantum state preparation framework that adopts a learning-based approach yet differs fundamentally from variational or hybrid QSP methods. Instead of repeatedly executing a parameterized circuit within a classical optimization loop, our method performs all learning entirely in the classical domain. Before any quantum operation, a classical routine trains the parameters of a diagonal Hamiltonian whose simulated evolution generates the target amplitudes.

Our approach builds directly on the Hamiltonian-simulation paradigm pioneered by Feynman, Lloyd, and subsequently formalized for sparse Hamiltonians by Berry, Childs, Low, and others \cite{feynman2018simulating, lloyd1996universal, berry2007efficient, berry2009black, childs2010simulating, berry2014exponential, low2017hamiltonian, low2017optimal, low2019hamiltonian}. These works establish that sparse Hamiltonians can be simulated efficiently when one has oracle access to their matrix elements; diagonal Hamiltonians constitute a particularly simple subclass, requiring only the ability to query the diagonal entries. We exploit exactly this property: once the Hamiltonian parameters have been classically learned, their oracle-based simulation enables the required evolution using a constant number of queries. Motivated by the millisecond-scale coherence limits of current superconducting and trapped-ion hardware, this off-device training eliminates the repeated quantum-evaluation cycles inherent to variational schemes and allows the entire state-preparation stage to be executed via a constant-depth quantum evolution.

Specifically, we demonstrate that the quantum time complexity can be reduced from $O(N)$ to $O(1)$ by incorporating a classical learning procedure of cost $O(N\log{N})$. Here, the $O(1)$ refers to the number of queries to an oracle that provides the coefficients of a diagonal Hamiltonian used in the simulation. Rather than relying on oracle-access to the dataset values themselves, the proposed scheme shifts oracle-access to the Hamiltonian parameters, allowing the quantum circuit to synthesize the target state through a constant number of diagonal evolutions. In the standard amplitude-oracle model, Grover's method achieves $O(\sqrt{N})$ scaling \cite{grover2000synthesis}, and Bausch's structured-oracle construction reduces this to $O(1)$ only for specific amplitude distributions \cite{bausch2022fast}. By contrast, leveraging oracle-access to the Hamiltonian terms enables our method to attain $O(1)$ quantum query complexity for arbitrary datasets, with all nontrivial computational cost absorbed into the classical preprocessing stage.

Although the classical preprocessing cost formally scales as $O(N\log{N})$, one should note that asymptotic complexity does not directly reflect the practical runtime on real hardware. First, the constant prefactor associated with the classical computation is extremely small: our numerical analysis shows that the effective runtime behaves as approximately $10^{-7}\times N\log{N}$, a consequence of the efficiency of classical arithmetic operations and optimized linear-algebra kernels. Thus, even though $O(N\log{N})$ is asymptotically larger than $O(N)$, the actual wall-clock time of the classical phase remains negligible compared with any nontrivial quantum evolution. Second, the classical preprocessing benefits significantly from parallelism. Modern CPUs and GPUs perform large-scale vectorized operations that modify the effective scaling for moderate system sizes. Empirically, our implementation exhibits a sublinear effective scaling close to $O(N^{1/4})$ on our current processor for problem sizes up to $N=2^{22}$, reflecting the benefit of hardware-level parallelism. For larger $N$, where this parallelism saturates, the runtime reverts to the expected $O(N\log{N})$ scaling. We note, however, that more advanced or highly parallel classical architectures may sustain the sublinear regime over larger problem sizes or even achieve effective scalings below $O(N^{1/4})$. Taken together, these observations show that the classical overhead is practically insignificant, while the reduction of quantum time complexity from $O(N)$ to $O(1)$ is critically important given the coherence-time limitations of current quantum processors.

This separation of roles between classical preprocessing and quantum execution illustrates a broader principle that we refer to as \textit{complexity transfer}: shifting the algorithm's asymptotically expensive components to classical computation, while keeping the quantum circuit depth constant and coherence-compatible. Because classical resources can exploit massive parallelism and operate without decoherence constraints, transferring the dominant workload off-device enables quantum algorithms whose remaining quantum component is both minimal and hardware-compatible. We view this mechanism as a general strategy for achieving practical quantum advantage in the Noisy Intermediate-Scale Quantum (NISQ) era, a paradigm we previously applied to reduce the space complexity of the traveling-salesman problem \cite{ramezani2024reducing}.

Furthermore, we show that for structured datasets, direct oracle access to the Hamiltonian can be eliminated entirely. Expanding the diagonal Hamiltonian in a Walsh basis \cite{welch2014efficient} and retaining only a polynomial number of coefficients enables state preparation in polynomial time $\text{poly}(n)$, without significant fidelity loss. By choosing Walsh operators aligned with native qubit connectivity (e.g., one-local and nearest-neighbor two-local interactions), this truncated model not only remains analytically tractable but can be implemented directly on hardware-constrained Quantum Processing Units (QPUs). The resulting formulation thus balances theoretical efficiency with experimental realizability, presenting a scalable and physically motivated route toward near-term quantum supremacy.

\section{Literature Review}

The study of QSP has long established that preparing an arbitrary $n$-qubit state requires resources scaling exponentially with $n$. Early structured approaches, such as the method of Grover and Rudolph, generated superpositions corresponding to integrable probability density functions using a ladder of uniformly controlled rotations, achieving circuit depth $O(n 2^n)$ \cite{GroverRudolph02}. 

Subsequent general-purpose methods employed recursive decompositions. The seminal work of M{\"o}tt{\"o}nen \emph{et al.} introduced an exact deterministic algorithm based on uniformly controlled rotations (UCRs), yielding depth $O(N)$ with $2^{n+2}-4n-4$ controlled-NOT (CNOT) gates~\cite{Mottonen05}, later refined by Bergholm \emph{et al.}, who reduced the CNOT count to ${2^{n+1}-2n-2}$~\cite{Bergholm05}. Broader synthesis frameworks, such as those of Shende \emph{et al.}, confirmed that any such exact preparation requires $O(N)$ gates in the worst case~\cite{Shende06}. Plesch and Brukner further optimized these decompositions using universal gate constructions, achieving an asymptotically improved CNOT count $\frac{23}{24} 2^n$ and roughly halving the depth~\cite{Plesch11}.

In parallel, several works used ancillary qubits to reduce depth. The Divide-and-Conquer State Preparation (DCSP) protocol of Araujo \emph{et al.} achieved depth $O(n^{2})$ with $O(N)$ qubits, though the ancilla qubits remain entangled~\cite{Araujo21}. Ultimately, Sun \emph{et al.} established that exact deterministic QSP can reach the theoretical minimum depth $\Theta(n)$ with $\Theta(N)$ ancillae \cite{Sun23}, and Gui \emph{et al.} showed that the SP+CSP architecture achieves optimal $\Theta(n)$ depth with $\Theta(N)$ total space-time cost \cite{Gui24}. Nmaju \emph{et al.} extended DCSP into a deterministic, measurement-based version with depth $O(n)$ using $O(N)$ ancillae \cite{nmaju2025low}.

Because exact preparation of arbitrary states is unavoidably exponential, later research focused on exploiting sparsity. In Sparse QSP (SQSP), where only $d$ amplitudes are nonzero, Gleinig and Hoefler achieved an exact scheme of size $O(dn)$ without ancillae \cite{Gleinig21}, and Malvetti \emph{et al.} developed a permutation/Householder-based approach with the same complexity \cite{Malvetti21}. Mozafari \emph{et al.} proposed a deterministic divide-and-distribute method reaching $O(kn)$ complexity using just one ancilla qubit, where $k\le d$ is the number of paths in the decision diagram \cite{Mozafari22DD}. Li and Luo later proved near-optimal bounds, showing SQSP without ancillae can be done in $O\left(\frac{nd}{\log n}+n\right)$ gates \cite{LiLuo25}.

The first oracle-based framework was proposed by Grover \cite{grover2000synthesis}, who proposed a quantum algorithm that exploits oracle access to the dataset to efficiently synthesize the target state. His method assumes the existence of a quantum oracle represented by the unitary operator
\begin{equation}
	U: \ket{j}\ket{0} \rightmapsto \ket{j}\ket{x_{j}},
\end{equation}
which returns the data value $x_j$ for a given index $j$. Under this oracle model, Grover showed that the desired state can be prepared with time complexity $O(\sqrt{N})$
using amplitude amplification---a significant improvement over the $O(N)$ deterministic bound. 

For each round of amplitude amplification, Grover's method requires $O(n^2)$ gates to perform arithmetic operations. Sanders \emph{et al.}~\cite{sanders2019black} improved upon this by replacing the arithmetic calculations with a simple inequality test that compares the oracle output against a uniform superposition of all possible values. This modification reduces the gate complexity per round to $O(n)$ while maintaining the same $O(\sqrt{N})$ overall time complexity, making the algorithm significantly more practical for implementation.

Whereas Grover and Sanders \emph{et al.} both achieve $O(\sqrt{N})$ time complexity, Bausch \cite{bausch2022fast} demonstrates that this bound can be significantly improved---even down to $O(1)$ in some cases---by exploiting prior knowledge about the amplitude distribution. The key insight is that the initial state in amplitude amplification can be optimized based on the average bit weight of the amplitudes. Specifically, Bausch replaces the uniform gradient state used in Sanders' approach with an optimized gradient state that reflects the expected bit configuration of the target amplitudes. For structured distributions such as uniform, triangular, or certain power-law distributions, this modification reduces the number of amplitude amplification rounds from $O(\sqrt{N})$ to $O(1)$, achieving an exponential speedup.

Moosa \cite{moosa2023linear} develops a Fourier-Series Loading (FSL) algorithm for quantum state preparation, where the target function is approximated using a truncated Fourier expansion containing $m$ terms. The circuit first encodes the discrete Fourier coefficients onto qubit amplitudes and then applies an inverse Quantum Fourier Transform (QFT) to obtain the state. Because the QFT has gate complexity $O(n^2)$ and depth $O(n)$, while the coefficient-loading subroutine scales with the number of retained Fourier modes, the total cost is $O(2^m + n^2)$ with depth $O(2^m + n)$. 

Zylberman \cite{zylberman2024efficient} introduces a probabilistic Walsh-Series Loading (WSL) method for quantum state preparation, in which a target function \(f(x)\) is approximated by a truncated Walsh expansion. The parameters \(\epsilon_0\) and \(\epsilon_1\) set the accuracy of two approximations: \(\epsilon_0\) controls the amplitude scaling precision and \(\epsilon_1\) determines the truncation order of the Walsh series, with smaller values yielding higher accuracy but greater circuit cost. In the one-dimensional case, the resulting circuit---composed of a superposition initialization followed by a controlled-diagonal unitary with \(M = 2^{\lfloor \log_2(1/\epsilon_1)\rfloor + 1}\) Walsh operators---achieves size \(O(n + 1/\sqrt{\epsilon_1})\) and depth \(O(1/\sqrt{\epsilon_1})\), succeeding with probability \(\Theta(\epsilon_0^2)\).

McArdle \emph{et al.} \cite{mcardle2022quantum} present a white-box, QSVT-based template for preparing states with amplitudes given by a known function $f$, eliminating the need for amplitude oracles, coherent reversible arithmetic, or QRAM. The method first block-encodes $\sin(.)$ on the computational basis via a phase-gradient Hadamard-test construction, then applies a definite-parity polynomial $h$ with $|h(y)|\le1$ on $[-1,1]$ so that $f(x)\approx h(\sin(x/a))$; QSVT implements $h$ to produce a block-encoding of the target amplitudes. The resource scaling is governed by a "discretized L2-norm filling-fraction" $\mathcal{F}$, reflecting how much L2-mass the function places relative to its peak, which enters both accuracy bounds and success behavior. The main theorem shows one can prepare a state $\epsilon$-close (in trace distance) to $\ket{\psi}$ using a template circuit with gate complexity proportional to $n d\mathcal{F}$ (where $d$ is the polynomial degree) and at most three ancilla qubits; there are no amplitude-oracle calls. For functions admitting exponential-quality polynomial approximations (smooth/analytic profiles), the required degree satisfies $d=O(\log(1/\epsilon))$, yielding near-linear scaling in $n$ with modest overhead from $\mathcal{F}$.

Gonzalez-Conde \cite{gonzalez2024efficient} develops an efficient QSVT-based framework for quantum amplitude encoding of polynomial functions, extending prior Walsh-series loading ideas where the linear function itself is realized through a Walsh-series expansion. The procedure first block-encodes this Walsh-based linear function on \(n\) qubits using a circuit of depth \(O(n)\), establishing the variable register in amplitude form. Then, a QSVT layer applies a sequence of controlled rotations determined by the coefficients of a target polynomial \(P(y)\), coherently transforming amplitudes \(y_j\) to \(P(y_j)\) and producing
\[
\ket{\psi_P} \propto \sum_j P(y_j)\ket{j}.
\]  
The overall complexity scales linearly with the number of qubits \(n\) and polynomially with the degree \(d\) of \(P(y)\). 

Zoufal \emph{et al.} \cite{zoufal2019quantum} proposed a variational, learning-based state preparation method using Quantum Generative Adversarial Networks (qGANs). The scheme trains a parameterized quantum generator-a layered $R_Y + CZ$ variational circuit of depth $k$-together with a classical discriminator to reproduce a target probability distribution from data samples. The trained generator prepares a quantum state
\[
\ket{\psi_\theta} \propto \sum_j \sqrt{p_j^{(\theta)}}\ket{j},
\]
approximating the desired distribution, with gate complexity $O(kn)$ for $n$ qubits. Implementations on simulators and IBM Q hardware were limited to three qubits, successfully loading log-normal, triangular, and bimodal distributions. The scalability is discussed only theoretically, assuming polynomial resource growth but lacking empirical evidence. The paper also notes that training convergence depends heavily on dataset structure---smooth or low-dimensional distributions train reliably, while random or unstructured data lead to unstable, nonconvergent cost functions.

Nakaji \cite{nakaji2022approximate} introduces a shallow, variational approach to approximate amplitude encoding, where a parameterized quantum circuit (PQC) is trained directly to reproduce a target amplitude distribution using a cost-minimization loop based on measurable statistical distances such as maximum mean discrepancy (MMD). This contrasts with Zoufal \emph{et al.}, which trains a PQC adversarially within a quantum-classical GAN framework. Nakaji's method eliminates the need for a classical discriminator, achieving stable, single-loop optimization with polynomial gate cost $O(kn)$ suitable for NISQ devices. Whereas both are approximate and data-driven, Nakaji's approach is deterministic in training and designed for function and time-series encoding (e.g., financial data), whereas Zoufal's qGAN is adversarial, focused on distribution learning and prone to convergence instability.

Marin-Sanchez \emph{et al.} \cite{marin2023quantum} extend function-loading algorithms by introducing a variational method that complements analytical Fourier-based schemes. The method trains a parameterized quantum circuit $U(\theta)$ to approximate the target function's amplitude profile through a fidelity-based cost function, executed in a hybrid classical-quantum loop. Unlike the adversarial qGAN (Zoufal) or unconstrained PQC fitting (Nakaji), this approach uses a structured, function-adaptive ansatz where parameters are clustered around features such as zeros and singularities, reducing redundancy and improving convergence. Only a few tunable parameters are needed, yielding low-depth circuits that accurately load irregular functions while avoiding barren plateaus, thereby making variational state preparation efficient even for nonsmooth or singular functions.

\section{Method}

\begin{figure}[htb]
\centering
\includegraphics[width=0.48\textwidth]{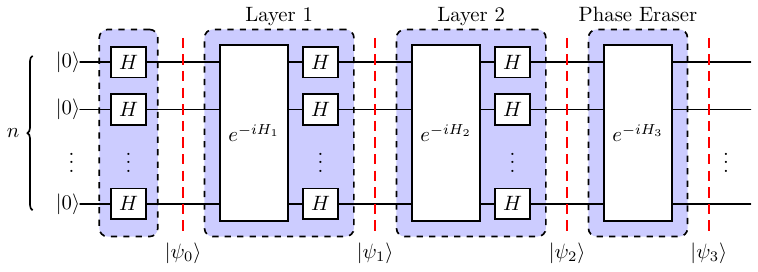}
\caption{Quantum circuit of the proposed Hamiltonian-simulation-based state preparation algorithm.}
\label{fig-qc}
\end{figure}

The quantum circuit underlying our state preparation algorithm is depicted in Fig. \ref{fig-qc}. The procedure is based on combining Hamiltonian simulation layers with interleaved Hadamard transformations, which together map phase modulation into amplitude encoding suitable for quantum state synthesis.

%\begin{figure}[htb]
%	\centering
%	\includegraphics[width=\columnwidth]{pic/fig-qc.pdf}
%	\caption{Quantum circuit of the proposed Hamiltonian-simulation-based state preparation algorithm.}
%	\label{fig-qc}
%\end{figure}

The circuit begins with a Hadamard layer applied to all qubits, generating the uniform superposition of computational basis states
\begin{equation}
	\ket{\psi_{0}}=\frac{1}{\sqrt{N}} \sum_{j=0}^{N-1}\ket{j}.
\end{equation}
Subsequently, two sequential layers of Hamiltonian evolutions are applied, each followed by another Hadamard layer. The Hamiltonians used in these evolutions are assumed to be diagonal in the computational basis, defined as
\begin{equation}\label{eq-hamiltonian}
	H_{k} = \sum_{j=0}^{N-1}h_{k,j}\ket{j}\bra{j}, \quad k\in\{1,2,3\},
\end{equation}
where $h_{k,j}$ are real-valued parameters. During the quantum evolution governed by each diagonal Hamiltonian, the basis state $\ket{j}$ acquires a phase factor $e^{-ih_{k,j}}$, so the Hadamard gate following each simulation converts these accumulated phases into amplitude distributions over the computational basis. After the second Hamiltonian layer, the state of the system can be expressed as
\begin{equation}
	\ket{\psi_{2}} = \sum_{j=0}^{N-1} y_{j}(\mathfrak{h}_{2}) e^{-i\theta_{j}(\mathfrak{h}_{2})}\ket{j},
\end{equation}
where both the amplitude $y_{j}(\mathfrak{h}_{2})$ and the residual phase $\theta_{j}(\mathfrak{h}_{2})$ depend on the Hamiltonian parameters ${\mathfrak{h}_{2}=\{h_{1,j},h_{2,j}\}}$. 

To achieve the desired target state amplitudes $x_{j}$, the parameters $h_{1,j}$ and $h_{2,j}$ are optimized in a classical training stage. The loss function guiding this optimization is the Sum-of-Squares Error (SSE) between obtained and target amplitudes:
\begin{equation}
	L=\sum_{j=0}^{N-1}|y_{j}(\mathfrak{h}_{2}) - x_{j}|^2.
\end{equation}

After the minimization process, the amplitudes $y_{j}$ converge to $x_{j}$. Denoting the trained Hamiltonian parameters by $h_{k,j}^{(t)}$, the remaining phases of the states are given by $\theta_{j}(\mathfrak{h}_{2}^{(t)})$. These residual phases are compensated by a final Hamiltonian evolution layer, implemented as
\begin{equation}
	H_{3} = \sum_{j=0}^{N-1}-\theta_{j}(\mathfrak{h}_{2}^{(t)})\ket{j}\bra{j}.
\end{equation}

Applying this third evolution cancels the accumulated phases, resulting in the final prepared state
\begin{equation}
	\ket{\psi_{\text{3}}} = \sum_{j=0}^{N-1} x_{j} \ket{j}.
\end{equation}

\begin{figure}[t]
\centering
\includegraphics[width=0.48\textwidth]{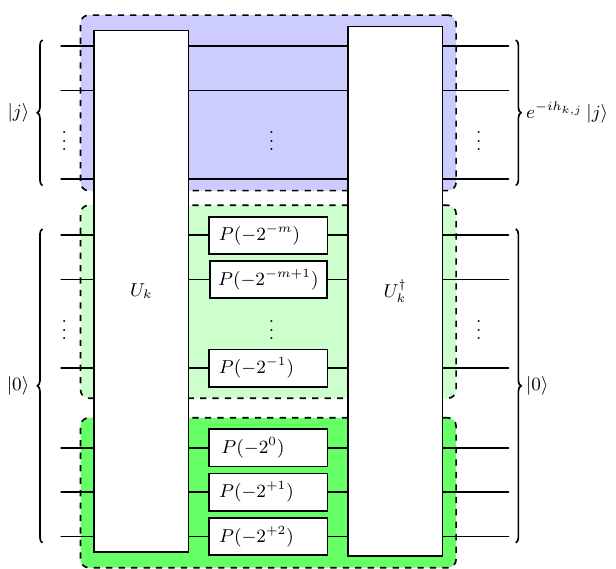}
\caption{Quantum circuit for simulating diagonal Hamiltonians using oracle access. The oracle $U_{k}$ encodes the coefficient $h_{k,j}$ into an ancillary register, single-qubit phase gates $P(\theta)$ apply the corresponding evolution $e^{-ih_{k,j}}$ to the computational basis state, and the inverse oracle $U_{k}^{\dagger}$ uncomputes the ancilla to $\ket{0}$}
\label{fig-HamSim}
\end{figure}

%\begin{figure}[t]%htb
%	\centering
%	\includegraphics[width=0.48\textwidth]{pic/fig-HamSim.pdf}
%	\caption{Quantum circuit for simulating diagonal Hamiltonians using oracle access. The oracle $U_{k}$ encodes the coefficient $h_{k,j}$ into an ancillary register, single-qubit phase gates $P(\theta)$ apply the corresponding evolution $e^{-ih_{k,j}}$ to the computational basis state, and the inverse oracle $U_{k}^{\dagger}$ uncomputes the ancilla to $\ket{0}$}
%	\label{fig-HamSim}
%\end{figure}

If oracle access to the Hamiltonian is available, the simulation of diagonal evolutions can be implemented efficiently \cite{childs2004quantum}. Specifically, for a given diagonal Hamiltonian $H_k$, we assume access to a unitary oracle $U_k$ such that
\begin{equation}
	U_k\ket{j}\ket{0}=\ket{j}\ket{h_{k,j}},
\end{equation}
where the second register encodes the binary representation of $h_{k,j}$ using $m+3$ qubits. Here, the 3 bits represent the integer part of $h_{k,j}\in[0,2\pi)$, and the $m$ bits provide a precision of $2^{-m}$. This size is sufficient because the Hamiltonian parameters can be defined modulo $2\pi$ due to the additive gauge freedom in the global phase, and the phase evolution $e^{-ih_{k,j}}$ is $2\pi$-periodic.

The corresponding circuit implementation is shown in Fig. \ref{fig-HamSim}. After the oracle $U_k$ is applied, the ancillary register contains a reversible encoding of $h_{k,j}$. A sequence of single-qubit phase gates $P(\theta)$ then acts directly on these encoded qubits according to $P(\theta)\ket{l}=e^{i\theta l}\ket{l}$ for $l\in\{0,1\}$, so the cumulative effect of all phase rotations is to imprint a global factor $e^{-ih_{k,j}}$ on the joint state $\ket{j}\ket{h_{k,j}}$. The oracle's inverse $U_{k}^{\dagger}$ is subsequently applied to uncompute the ancillary register, restoring it to $\ket{0}$ while leaving the computational register $\ket{j}$ with the acquired phase $e^{-ih_{k,j}}$.

The classical simulation of the circuit, which we use throughout the learning phase to optimize the Hamiltonian parameters, is particularly efficient because both components of each layer admit fast classical implementations. The global Hadamard gates are evaluated using the Fast Walsh-Hadamard Transform (FWHT), yielding an $O(N\log{N})$ update instead of a dense $O(N^2)$ matrix multiplication. Meanwhile, the diagonal Hamiltonian evolution $e^{-iH_{k}}$ requires no matrix operations at all: its action reduces to elementwise multiplication of the form $\psi_{j}\rightarrow e^{-ih_{k,j}}\psi_{j}$ which runs in linear time $O(N)$. 

Importantly, this structure also reduces the space complexity of the simulation. In a general classical simulation of a quantum circuit, representing an arbitrary $n$-qubit gate requires storing a full $N\times N$ matrix, resulting in $O(N^2)$ memory usage. By contrast, the FWHT operates entirely in place using only the state vector itself (plus $O(1)$ auxiliary scalars), and the diagonal evolution operator is stored as an $N$-dimensional vector of phases. Consequently, the entire classical simulation requires only $O(N)$ memory, matching the size of the state vector and avoiding the quadratic storage overhead inherent to arbitrary gate matrices.

As a result, each simulation step consisting of a Hadamard layer followed by a diagonal evolution is dominated by the FWHT, giving total runtime $O(N\log{N})$ and space complexity $O(N)$.

\begin{figure}[t]
\centering
\includegraphics[width=0.35\textwidth]{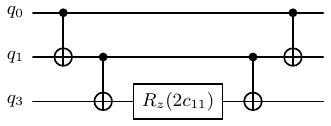}
\caption{Quantum circuit implementing the unitary $e^{-ic_{11}W_{11}}$ corresponding to the Walsh operator ${W_{11}=Z^{(3)}\otimes Z^{(1)}\otimes Z^{(0)}}$. Because the binary representation of $r=11$ is $1011$, the operation acts nontrivially on qubits 0, 1, and 3. A more depth-efficient decomposition achieving the same gate count can be found in \cite{sriluckshmy2023optimal}.}
\label{fig-r11}
\end{figure}

%\begin{figure}[t]%htb
%	\centering
%	\includegraphics[width=0.35\textwidth]{pic/fig-r11.pdf}
%	\caption{Quantum circuit implementing the unitary $e^{-ic_{11}W_{11}}$ corresponding to the Walsh operator ${W_{11}=Z^{(3)}\otimes Z^{(1)}\otimes Z^{(0)}}$. Because the binary representation of $r=11$ is $1011$, the operation acts nontrivially on qubits 0, 1, and 3. A more depth-efficient decomposition achieving the same gate count can be found in \cite{sriluckshmy2023optimal}.}
%	\label{fig-r11}
%\end{figure}

In many practical situations---particularly when the target data exhibit structure or smooth spectral correlations---the diagonal Hamiltonian can be compactly represented by a small number of terms in the Walsh operator basis. The Walsh basis forms a complete set of mutually commuting diagonal operators on $n$ qubits, defined as
\begin{equation}
	W_r = \sum_{j=0}^{N-1}(-1)^{r.j}\ket{j}\bra{j},
\end{equation}
where $r.j$ denotes the bitwise inner product of the binary strings corresponding to $r$ and $j$. Any diagonal Hamiltonian $H=\sum_{j} h_j \ket{j}\bra{j}$ can therefore be expressed as
\begin{equation}
	H = \sum_{r} c_{r} W_{r}, \quad c_r = \frac{1}{N} \sum_{j} (-1)^{r.j} h_{j}.
\end{equation}

When the dataset or target distribution possesses an inherent regularity, only a small subset ${\mathcal{R}\subseteq\{0,1,2,\dots,N-1\}}$ of coefficients contribute significantly. Truncating the expansion to these dominant components yields
\begin{equation}
	H_{\text{trunc}} = \sum_{r\in\mathcal{R}} c_{r} W_{r},
\end{equation}
reducing the description of the Hamiltonian to a compact, data-adapted form.

Because all Walsh operators commute and each ${W_r=Z^{r_{n-1}}\otimes Z^{r_{n-1}}\otimes \dots \otimes Z^{r_{0}}}$ is a tensor product of Pauli-$Z$ operators, the corresponding unitaries $e^{-ic_{r}W_{r}}$ decompose efficiently into single-qubit rotations and CNOT gates (Fig. \ref{fig-r11}). This commutativity also allows implementing all terms without Trotter error. In this formulation, the coefficients $c_r$ serve as trainable parameters within the hybrid optimization loop, enabling the algorithm to learn an effective truncated Hamiltonian directly from data.

When only a polynomial number of Walsh terms is selected, the corresponding diagonal Hamiltonian can be implemented on the quantum device with $\text{poly}(n)$ gate complexity using its explicit gate-level decomposition shown earlier, yielding a fully oracle-free realization of the truncated evolution. Although the classical simulation procedure is identical in both regimes, their real quantum execution differs fundamentally. In the oracle-based regime, once the amplitudes have been classically learned, the residual phases can be cancelled directly on the QPU through a final oracle-generated diagonal evolution, so no training of the last Hamiltonian is required. In the Walsh-truncated, oracle-free regime, such phase cancellation is not available during execution, and therefore the final diagonal evolution must itself be learned classically. Denoting trainable parameters by
\begin{equation}
	\mathfrak{c}_{3}=\{c_{1,r},c_{2,r},c_{3,r}\},
\end{equation}
the resulting state after third Hamiltonian evolution is
\begin{equation}
	\ket{\psi_{3}} = \sum_{j=0}^{N-1} z_{j}(\mathfrak{c}_{3}) e^{-i\vartheta_{j}(\mathfrak{c}_{3})}.
\end{equation}
To ensure that the amplitudes converge to the dataset while the unwanted phases are removed, we minimize
\begin{equation}
	L=\sum_{j=0}^{N-1}|z_{j}(\mathfrak{c}_{3}) - x_{j}|^2 + |\vartheta_{j}(\mathfrak{c}_{3})|^{2}.
\end{equation}
Thus, in the oracle-free Walsh-basis construction, both amplitude synthesis and phase erasure are performed entirely within the classical learning phase, while the final quantum execution uses only the explicitly decomposed $\text{poly}(n)$-sized diagonal evolution.

\section{Results}

\begin{figure*}[t]
	\centering
	\subfloat[Training loss vs.\ epoch (1 layer).]{%
		\includegraphics[width=0.48\textwidth]{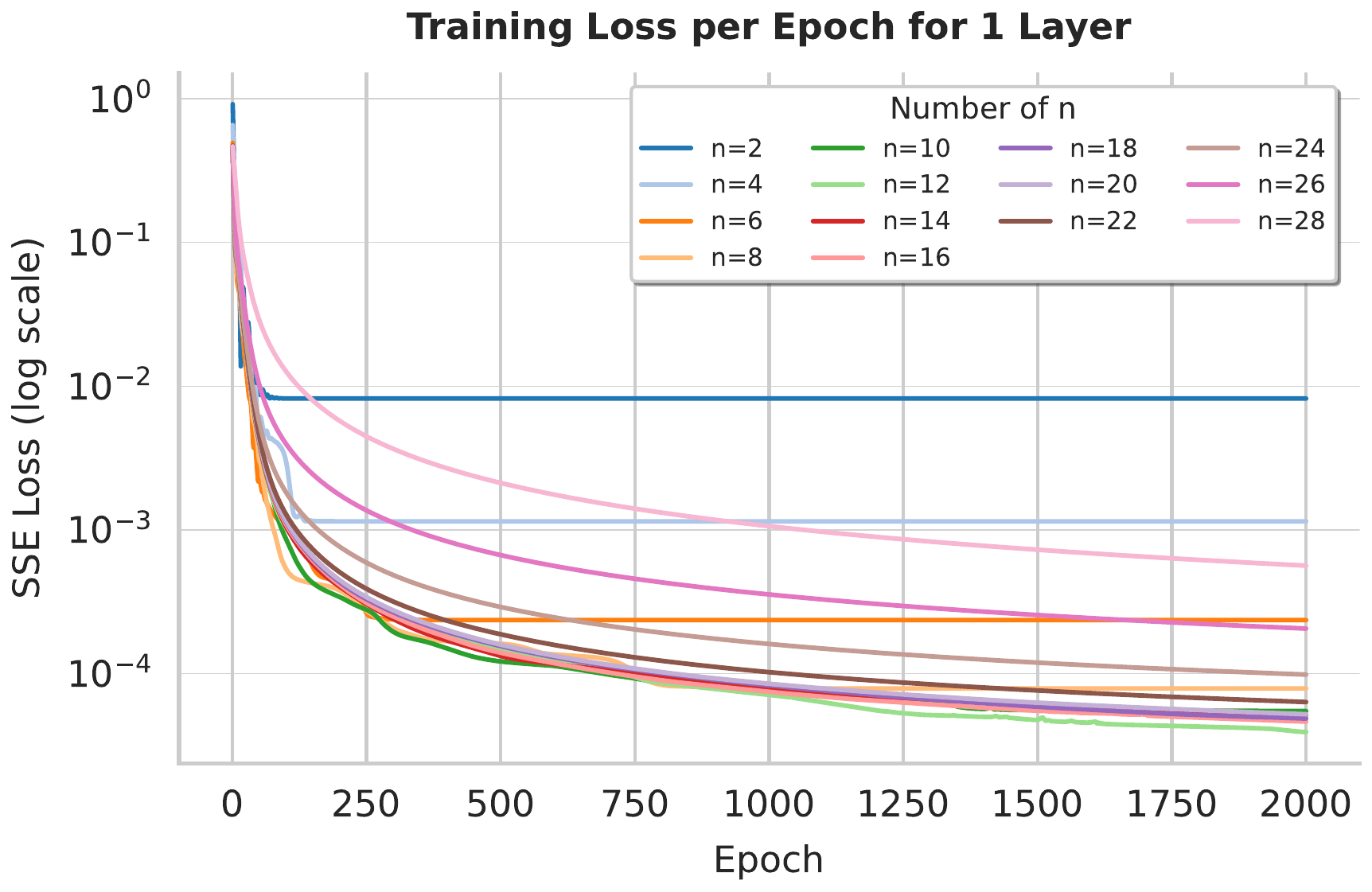}}
	\hfill
	\subfloat[Training loss vs.\ epoch (2 layers).]{%
		\includegraphics[width=0.48\textwidth]{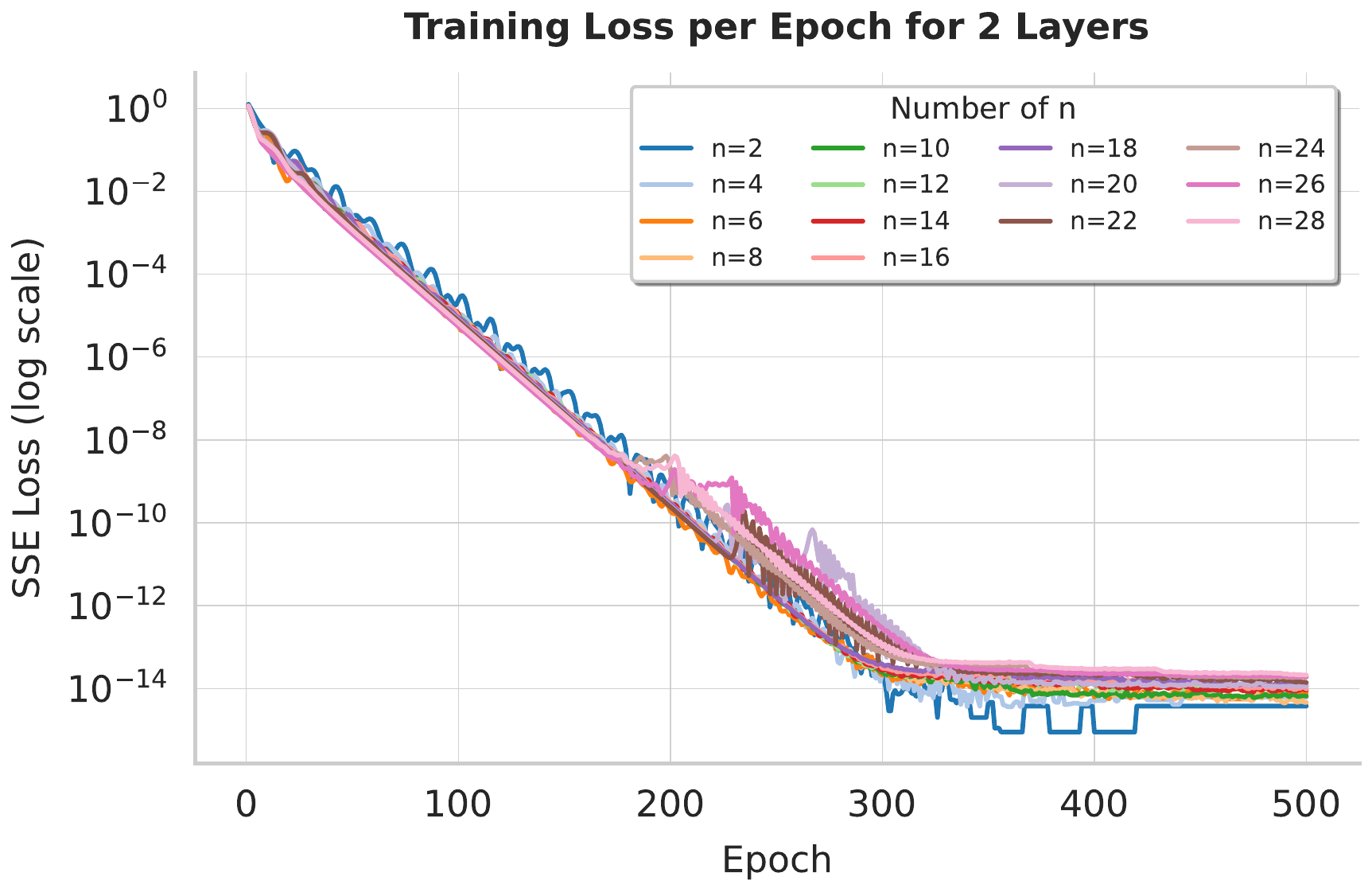}}
	
	\subfloat[Final loss vs.\ dataset size for uniform and normal samples.]{%
		\includegraphics[width=0.48\textwidth]{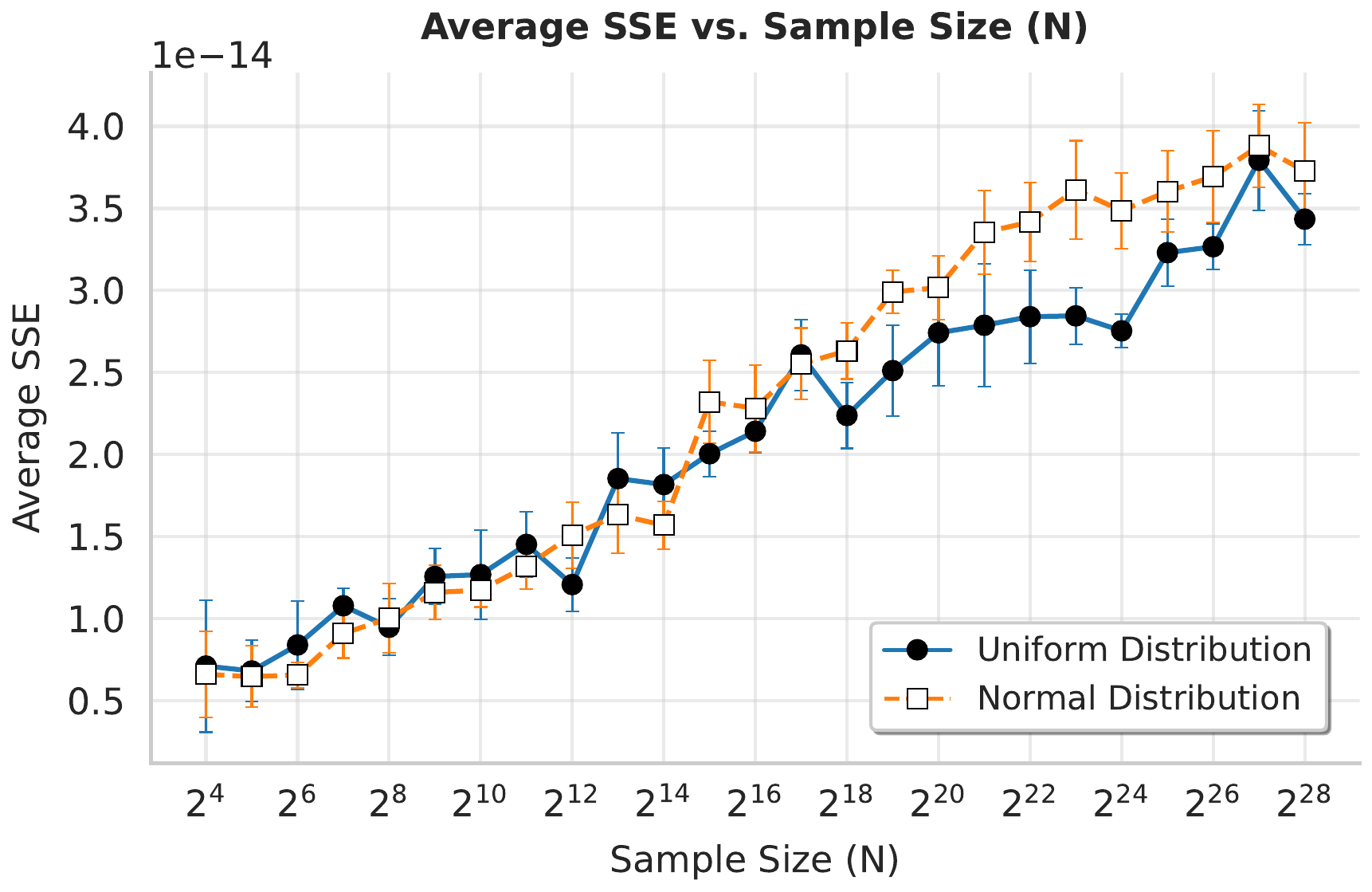}}
	\hfill
	\subfloat[Runtime scaling with dataset size.]{%
		\includegraphics[width=0.48\textwidth]{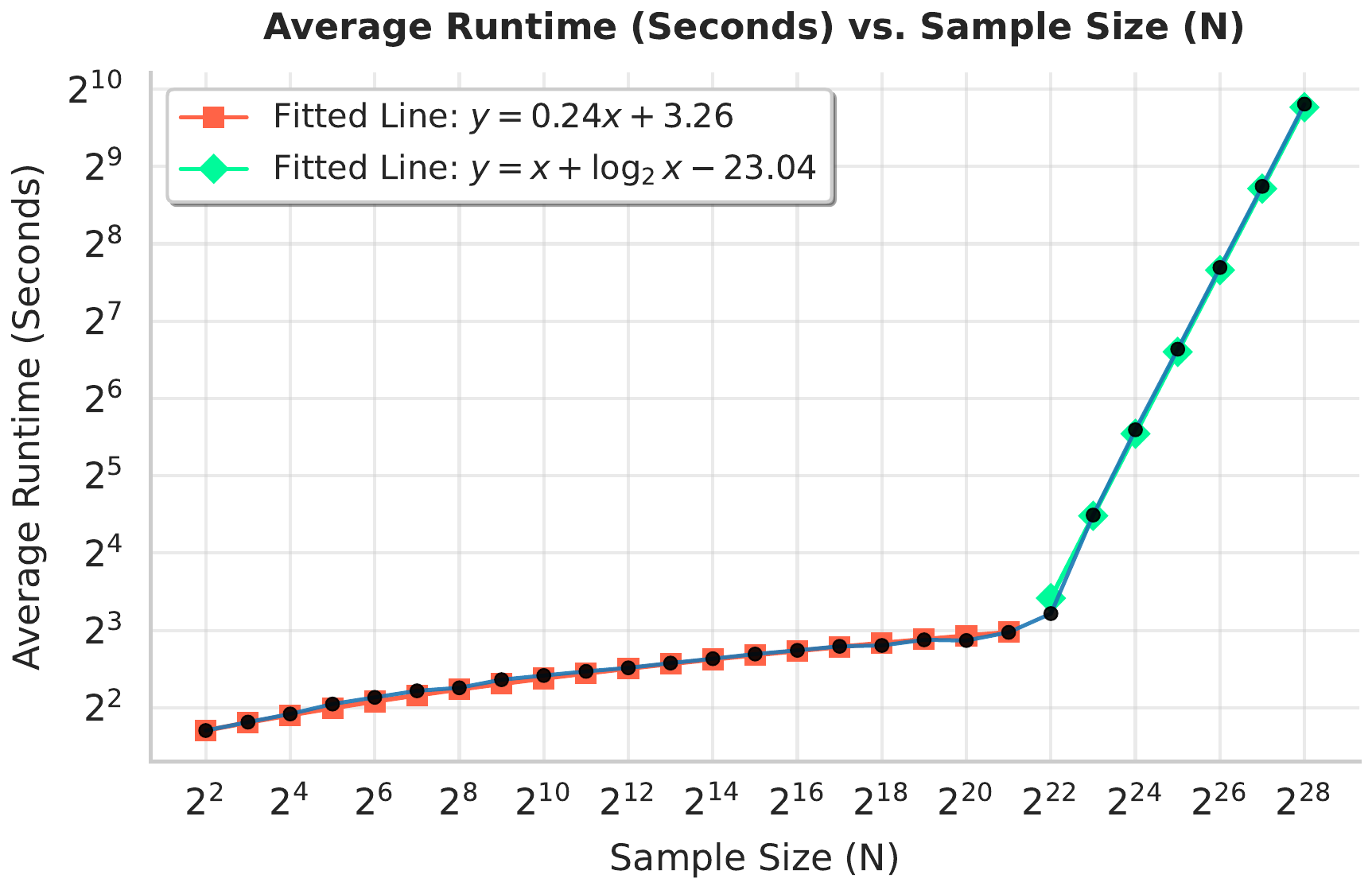}}
	\caption{Summary of training performance, sample-size dependence, and classical runtime scaling.}
	\label{fig-results}
\end{figure*}

In this section we report the performance of the proposed Hamiltonian-based state preparation approach. As described in the Method section, we introduce two versions of the algorithm. The first assumes oracle access to the diagonal Hamiltonian and trains all diagonal entries directly in the computational basis (Method 1). The second expands the Hamiltonian in the Walsh basis and trains only a selected subset of terms, enabling polynomial-size parametrizations and hardware-efficient implementations (Method 2). A common feature of both methods is the classical training phase, in which all trainable parameters of the Hamiltonian are optimized using gradient-based updates prior to quantum execution. We therefore begin by presenting the classical-training behavior of Method 1, and subsequently report the results obtained for Method 2 on structured datasets and hardware-efficient architectures.

We first analyze Method 1, in which the full set of $N$ diagonal entries of the Hamiltonian are treated as trainable parameters. All results in Fig. \ref{fig-results}(a)--(d) correspond to this setting. To assess the expressive capacity of the architecture and motivate our final circuit design, we compared the training performance of a single diagonal-evolution layer against a two-layer variant. Fig. \ref{fig-results}(a) and Fig. \ref{fig-results}(b) show the optimization behavior for both settings. For each system size, the training dataset consists of real values sampled uniformly from $[0,1]$, and normalized to define a target quantum state. The single-layer baseline reduces the loss only to approximately $10^{-4}$ after 2000 epochs and does not converge further, whereas the two-layer configuration exhibits rapid exponential-like decay of the loss and reaches values near $10^{-14}$ after roughly 350 epochs. This comparison indicates that a single diagonal evolution is insufficient to capture the full amplitude structure, while two successive diagonal evolutions separated by Hadamard layers provide the expressivity required for accurate state synthesis. Accordingly, we adopt the two-layer architecture for all remaining experiments.

Fig. \ref{fig-results}(c) investigates how the final loss depends on dataset size. For each size we train for 500 epochs and average the result over ten independent datasets. We consider both uniformly distributed samples and samples drawn from the standard normal distribution $\mathcal{N}(0,1)$. The Gaussian case is especially relevant because, in oracle-based loading techniques, this distribution still requires $O(\sqrt{N})$ queries even in settings where some structured distributions permit $O(1)$ complexity \cite{bausch2022fast}. Our results show that the proposed approach maintains small final loss for both distributions using only a constant number of Hamiltonian simulations, indicating that the optimization dynamics do not depend on such oracle-specific lower bounds.

Fig. \ref{fig-results}(d) reports the classical runtime for 500 epochs as a function of $N$. For system sizes up to $N = 2^{22}$ the runtime follows an empirical scaling of $O(9.58\,N^{0.24})$, which is explained by parallel utilization of GPU cores. For larger systems the runtime crosses over to $O(1.16 \times 10^{-7}\,N \log_2 N)$, consistent with the asymptotic cost of the FWHT.

\begin{figure*}[t]
	\centering
	\subfloat[Linear dataset: amplitude fidelity using Two-Local Hamiltonian]{%
		\includegraphics[width=0.50\textwidth]{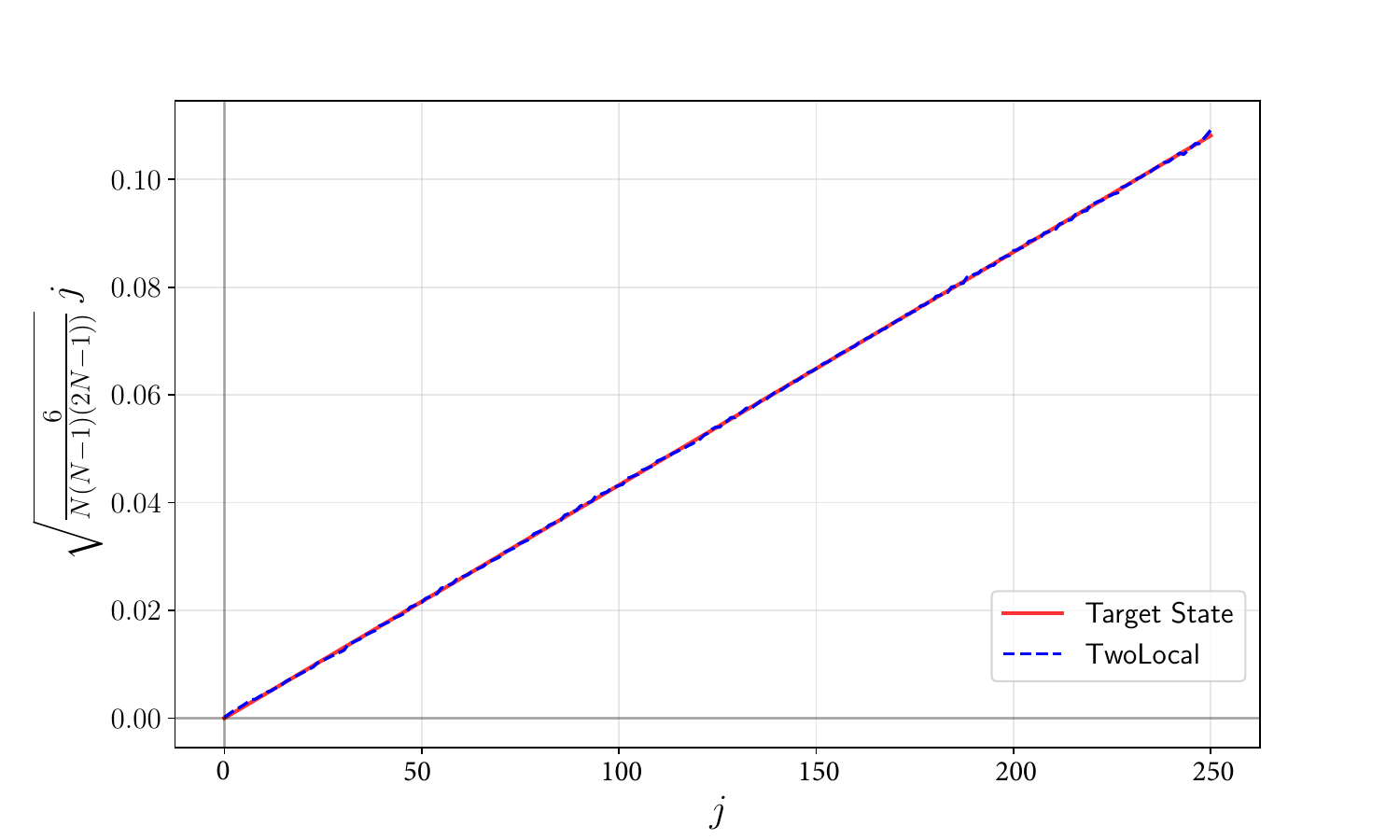}}
	\hfill
	\subfloat[Linear dataset: amplitude fidelity using Hardware-Efficient Hamiltonian]{%
		\includegraphics[width=0.50\textwidth]{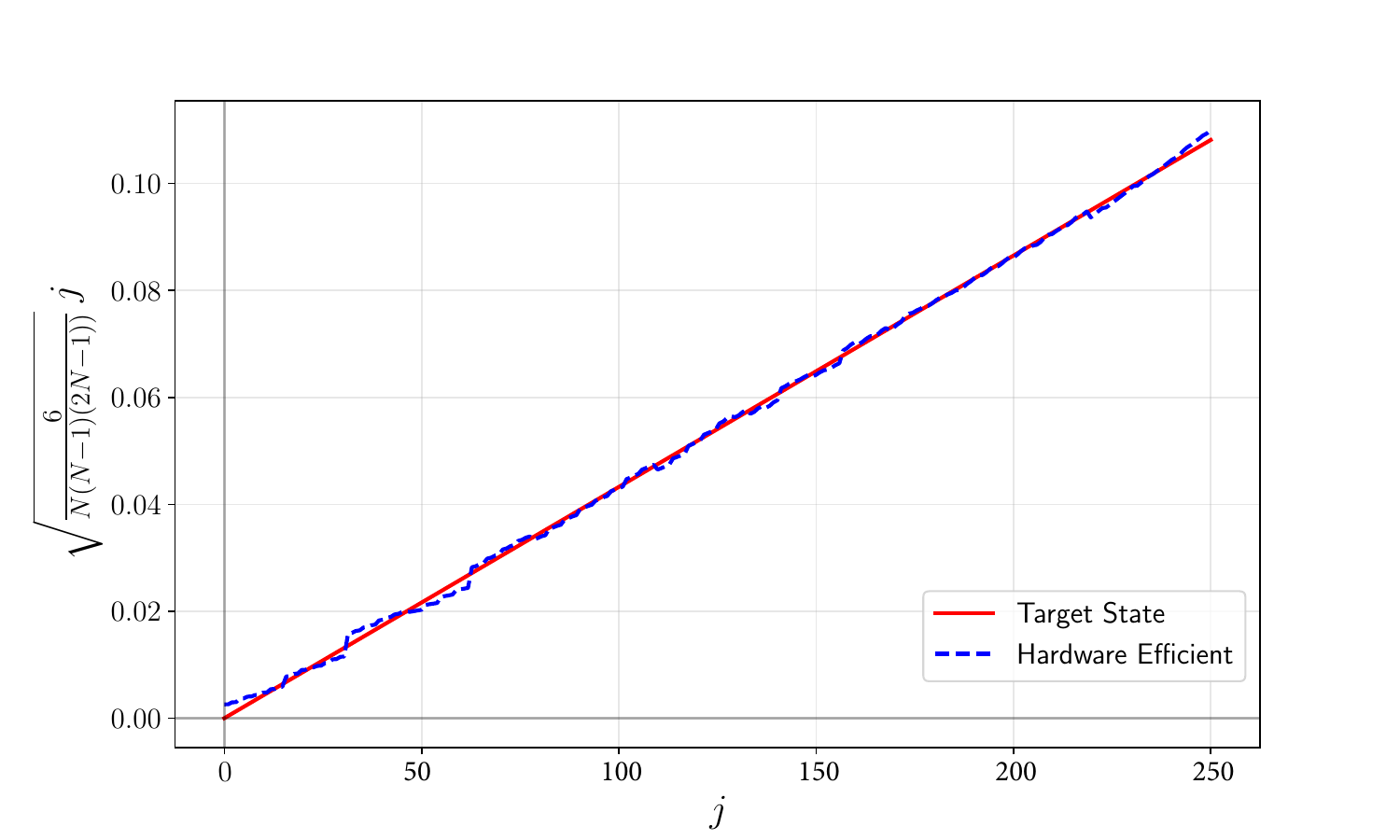}}
	
	\subfloat[Sine dataset: amplitude fidelity using Two-Local Hamiltonian]{%
		\includegraphics[width=0.50\textwidth]{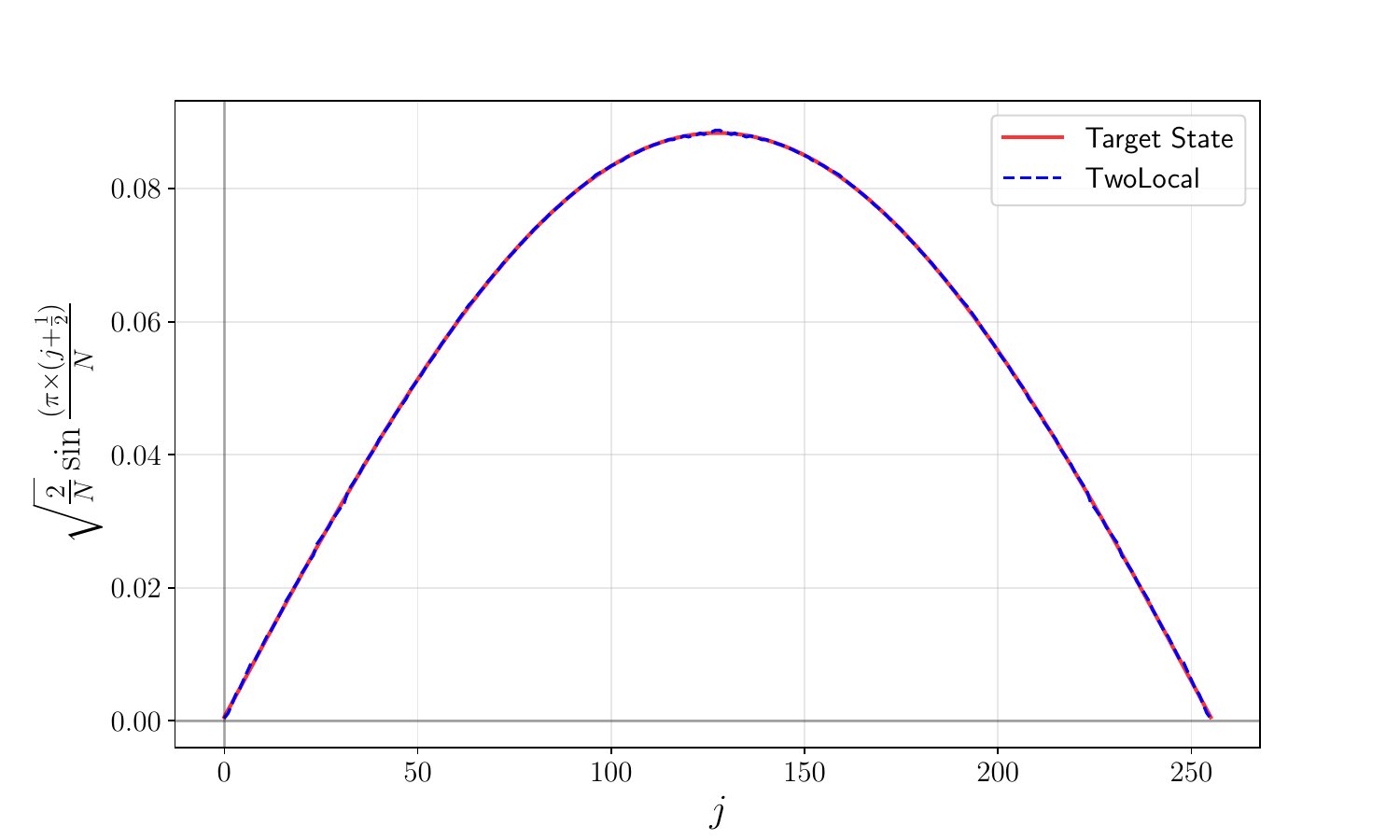}}
	\hfill
	\subfloat[Sine dataset: amplitude fidelity using Hardware-Efficient Hamiltonian]{%
		\includegraphics[width=0.50\textwidth]{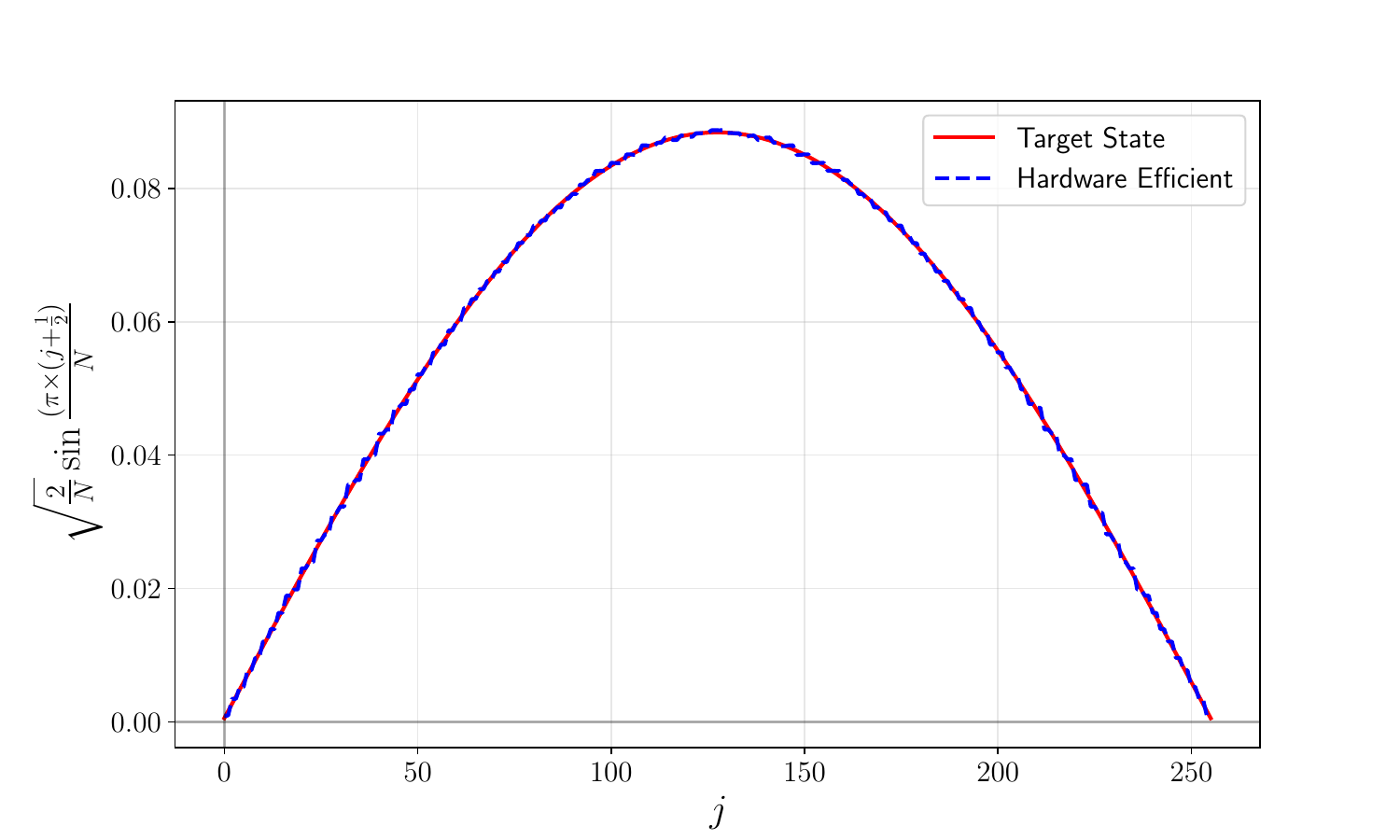}}
	\caption{Simulation results for structured datasets. Panels (a)--(d) show linear and sine state preparations using Two-Local and Hardware-Efficient Hamiltonians, illustrating amplitude reconstruction and fidelity comparison.}
	\label{fig-results-2}
\end{figure*}

\begin{figure}[htb]
	\centering
	\includegraphics[width=0.35\textwidth]{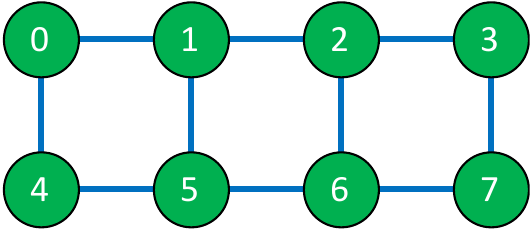}
	\caption{Ladder-topology connectivity of the qubit layout used in the hardware-efficient simulations}
	\label{fig-topology}
\end{figure}

We next evaluate Method 2, which targets structured datasets amenable to compact Walsh-basis representations. Two representative datasets are considered:
\begin{align}
	\ket{\phi_{1}} &= \sqrt{\frac{6}{N(N-1)(2N-1)}} \sum_{j=0}^{N-1} j \ket{j},\\
	\ket{\phi_{2}} &= \sqrt{\frac{2}{N}} \sum_{j=0}^{N-1} \sin{\frac{\pi(j+1/2)}{N}}\ket{j}.
\end{align}

The first state $\ket{\phi_{1}}$ corresponds to a linear amplitude profile. Linear data distributions are particularly relevant because, as shown by Gonz\'lez-Conde \emph{et al.} \cite{gonzalez2024efficient}, a properly normalized linear superposition can serve as an intermediate resource for function loading via QSVT, thereby enabling universal function synthesis from a single primitive state.

The second state $\ket{\phi_{2}}$, whose sinusoidal amplitude profile is closely connected to optimal phase estimation. As shown explicitly by Childs \cite{childs2010relationship}, sine-shaped initial states minimize the variance of finite-dimensional phase estimation procedures and thus achieve the tightest possible error bounds. Accurate phase estimation is also the central bottleneck in quantum linear-systems algorithms such as HHL, where eigenvalues obtained from phase estimation are subsequently inverted, leading to amplification of estimation errors proportional to the condition number. Consequently, sine-profile states play a distinguished role in both Hamiltonian-simulation-based quantum walks and error-mitigated subroutines of HHL-type algorithms, where variance minimization directly translates to improved numerical stability and reduced sensitivity to small eigenvalues.

For each dataset we restrict the trainable Hamiltonian terms to one-local and two-local Walsh components, where an $p$-local term refers to a basis index $r$ whose binary representation contains $p$ ones in the corresponding Walsh operator $W_{r}$. These correspond to the one-body $Z$ and two-body $ZZ$ interactions in the physical Hamiltonian. Two configurations are studied.

In the Two-Local model, all one- and two-local terms across the register are retained. In the Hardware-Efficient variant, only one-local and nearest-neighbor two-local terms are kept, consistent with the ladder-shaped QPU topology shown in Fig. \ref{fig-topology}. 

All simulations are performed with $N=2^{8}=256$ basis states. Fig. \ref{fig-results-2} illustrates the convergence behavior and the reconstructed amplitude distributions for both datasets.

Gate counts for Hamiltonian evolution are summarized in Table \ref{tab:gate_counts}, and the achieved state-preparation infidelities $(1-F)$ are summarized in Table \ref{tab:infidelity_results}.

\begin{table}[h]
	\centering
	\caption{Gate counts per diagonal-evolution layer}
	\begin{tabular}{lcc}
		\toprule
		\textbf{Configuration} & \textbf{One-qubit gates} & \textbf{Two-qubit gates} \\
		\midrule
		Two-Local & 36 & 56 \\
		Hardware-Efficient & 18 & 20 \\
		\bottomrule
	\end{tabular}
	\label{tab:gate_counts}
\end{table}

\begin{table}[h]
	\centering
	\caption{Infidelity of quantum state preparation ($1-F$)}
	\begin{tabular*}{\columnwidth}{@{\extracolsep{\fill}}lcc}
		\toprule
		\textbf{Dataset} & \textbf{Two-Local} & \textbf{Hardware-Efficient} \\
		\midrule
		Linear & $2.2\times10^{-5}$ & $3.3\times10^{-4}$ \\
		Sine   & $2.0\times10^{-5}$ & $1.1\times10^{-4}$ \\
		\bottomrule
	\end{tabular*}
	\label{tab:infidelity_results}
\end{table}

These results confirm that even with severe truncation of the Walsh expansion---restricted to physically local operators---the method preserves high preparation fidelity. Moreover, adapting the Hamiltonian to QPU connectivity yields circuits with nearly $3\times$ fewer gates while maintaining infidelity below $10^{-3}$, underscoring the scalability and hardware-efficiency of the proposed approach for structured data.

The gate counts in Table \ref{tab:gate_counts} are reported per diagonal-evolution layer. The following ablation instead counts CNOTs across the complete three-evolution circuit, so its resource axis is directly comparable across all retained Walsh supports.

\subsection{Walsh-support size and CNOT trade-off}

\begin{figure*}[t]
	\centering
	\includegraphics[width=\textwidth]{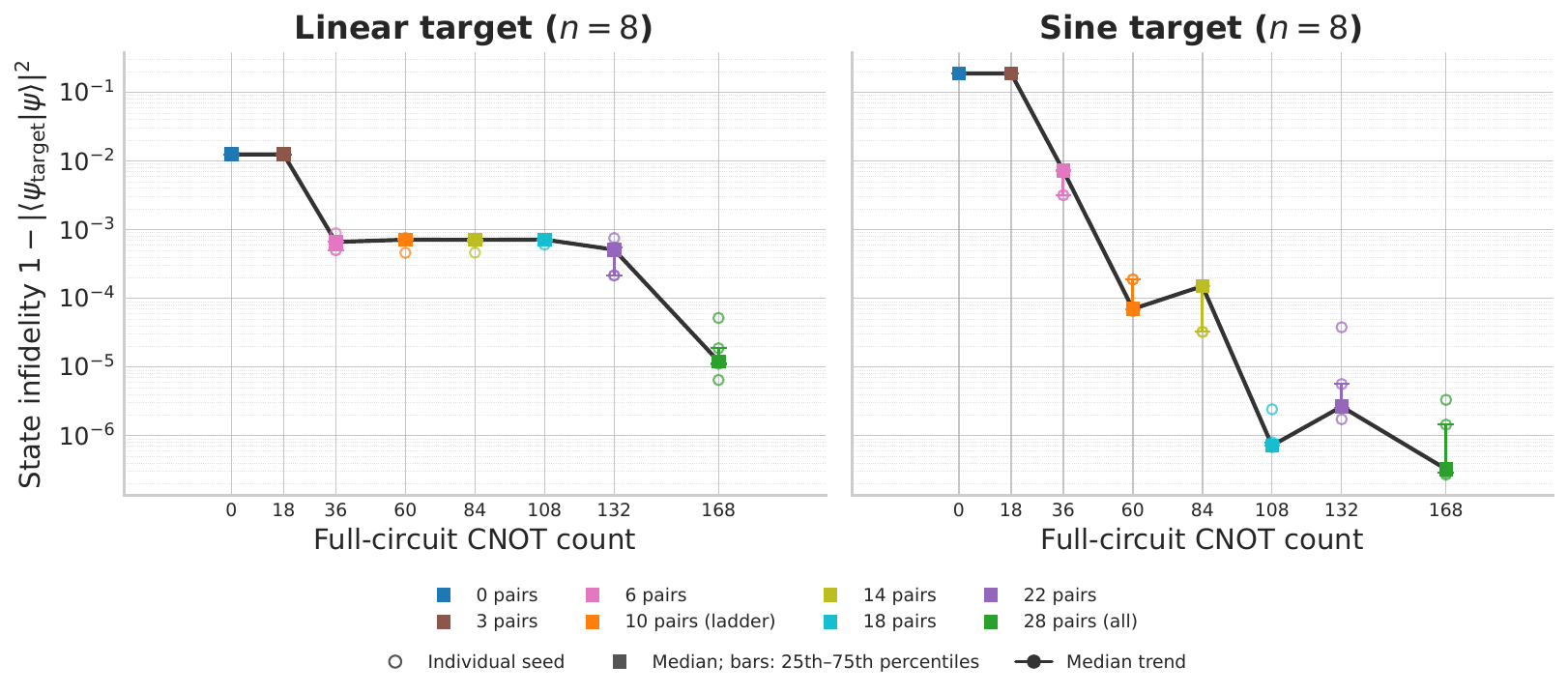}
	\caption{Accuracy--resource trade-off for nested Walsh supports at $n=8$. Open circles show individual seeds, squares and bars show medians and interquartile ranges, and the dark line connects support medians. The horizontal axis counts CNOTs across all three diagonal evolutions.}
	\label{fig-walsh-support-sweep}
\end{figure*}

To quantify how the retained Walsh support affects the accuracy--resource trade-off, we performed an additional $n=8$ ablation using a fixed nested support path. Starting from the one-local Hamiltonian, we added two-local Walsh terms in a deterministic order that first traversed the ten ladder edges and then added the remaining pair terms. This produced supports containing 0, 3, 6, 10, 14, 18, 22, and 28 pair terms, corresponding to full-circuit CNOT counts of 0, 18, 36, 60, 84, 108, 132, and 168, respectively. Each support was trained for both the linear and sine targets using five matched initializations and 2000 update epochs. For this resource ablation, we optimized and evaluated the normalized state infidelity; this protocol is separate from the earlier SSE-based simulations reported in Table \ref{tab:infidelity_results}.

The sweep shown in Fig. \ref{fig-walsh-support-sweep} shows that adding CNOTs generally improves accuracy, but the median curve is not strictly monotonic. This does not mean that a larger support is less expressive. Each support is trained separately, starting from random parameters and using a finite number of optimization steps. Training can therefore stop at different quality levels, so a larger support may give a worse final result than a smaller one. In principle, however, every larger support can reproduce the preceding support by setting its additional coefficients to zero. For the linear target, median infidelity decreases from $1.25\times10^{-2}$ at zero CNOTs to $7.16\times10^{-4}$ for the 60-CNOT ladder and $1.21\times10^{-5}$ for the all-pairs circuit. For the sine target, the corresponding values are $1.89\times10^{-1}$, $7.00\times10^{-5}$, and $3.25\times10^{-7}$. CNOT count is therefore a useful measure of circuit cost, but it is not the only factor that determines the accuracy reached by training. The experiment follows one fixed nested support path and is not an exhaustive search over all topologies with the same CNOT count.

\section{Conclusion}

In this work, we introduced a Hamiltonian-based framework for quantum state preparation (QSP) that fundamentally shifts computational effort from the quantum device to a classical preprocessing stage. The target quantum state is generated by simulating the evolution of a diagonal Hamiltonian whose parameters are learned classically, enabling the quantum circuit to remain shallow and coherence-compatible.

Once the Hamiltonian parameters are obtained, and assuming oracle access to these parameters, the desired quantum state can be prepared using only $O(1)$ queries to the Hamiltonian oracle, independent of the dataset size. Importantly, this constant query complexity is data-independent: the number of oracle calls does not depend on the structure, distribution, or statistics of the target amplitudes. This sharply contrasts with existing oracle-based approaches, most notably Bausch's structured-oracle construction, where $O(1)$ quantum complexity is attainable only for specific amplitude distributions and relies explicitly on prior knowledge of data structure. By transferring oracle access from the dataset itself to the Hamiltonian parameters, our method achieves $O(1)$ quantum query complexity for arbitrary datasets.

The cost of this generality is absorbed entirely by a classical learning phase with complexity $O(N\log{N})$, while the quantum execution remains constant-depth. This mechanism exemplifies what we refer to as complexity transfer: the asymptotically expensive portion of the algorithm is offloaded to classical computation, which is unconstrained by decoherence and can exploit parallel hardware. Because quantum state preparation is a subroutine in many quantum algorithms, reducing its quantum runtime is particularly significant in practice, where limited coherence times impose strict depth constraints on near-term quantum processors.

Furthermore, for structured datasets, our framework allows oracle access to be removed entirely. By expressing the Hamiltonian in the Walsh basis and retaining only a polynomial number of significant terms, high-fidelity state preparation can be achieved in $\text{poly}(n)$ time, where $n=\log_{2}{N}$. This demonstrates that intrinsic data structure can be systematically exploited to further reduce both quantum and classical overheads.

Finally, the proposed approach is naturally hardware-efficient. Constraining the Hamiltonian to one-local and two-local terms, or selecting parameters compatible with a given device topology, yields shallow circuits with reduced gate counts while preserving preparation fidelity. These results indicate that the method is not only asymptotically efficient, but also well aligned with the limitations of current and near-term quantum hardware.

\bibliographystyle{apsrev4-2}
\bibliography{references}

\end{document}